# Agile Free-Form Signal Filtering with a Chaotic-Cavity-Backed Non-Local Programmable Metasurface

Fabian T. Faul[1], Laurent Cronier[1], Ali Alhulaymi[2], A. Douglas Stone[2], Philipp del Hougne[1]*

[1] Univ Rennes, CNRS, IETR-UMR 6164, F-35000 Rennes, France

[2] Department of Applied Physics, Yale University, New Haven, CT 06520, USA

* Correspondence: philipp.del-hougne@univ-rennes.fr

**Filter synthesis is an inverse problem that is traditionally approached rationally by considering *spatially disjoint* resonators, approximating them as lumped elements, and engineering the coupling of selected pairs. This approach strongly limits the design space, making it challenging to build extremely tunable filters. Here, we demonstrate agile free-form signal filtering with an alternative purely-optimization-based design paradigm using a programmable system with many *spatially overlapping* modes. We back a programmable metallic metasurface with a quasi-2D chaotic cavity, inducing strong non-local interactions between all meta-elements and the connected ports. Thereby, the metasurface efficiently controls the transfer function between the ports. Our all-metallic device has unique advantages: ultra-wideband (UWB) tunability (7.5-13.5GHz), low loss, compactness, guaranteed linearity under high signal-power levels. First, we experimentally confirm theoretical predictions about reflectionless and transmissionless scattering modes; we also experimentally observe transmissionless exceptional points. Second, we impose diverse types of transfer function zeros at desired frequencies within an UWB range. Third, we achieve low-loss reflectionless programmable signal routing. Fourth, we investigate the trade-off between routing fidelity and bandwidth, achieving 20dB discrimination over 10MHz bandwidth. Fifth, we demonstrate UWB tunable multi-band filters that reject (<-24dB) or pass (≥-1dB) signals in specified bands whose centers, widths and number are reprogrammable.**

# 1. Introduction

Analog filtering is a ubiquitous need whenever electromagnetic signals are generated (to suppress out-of-band emissions) or received (to prevent receiver desensitization due to out-of-band stray signals). In fact, any wave-matter interaction can be interpreted as filtering the wave with the transfer function of the matter. Hence, antennas, signal routers and analog wave-based signal processors are all analog filters in a broader sense. To synthesize a filter with a desired transfer function, one is confronted with an inverse-design problem that is notoriously ill-posed. Microwave engineers have developed a suite of rational filter synthesis tools that only require some fine-tuning of the designed system during design closure to yield a *static* filter closely approximating the desired transfer function in a chosen bandwidth. However, as detailed below, these rational approaches severely limit the design space at the outset, making it very challenging to achieve *agile free-form* filtering, in which the transfer function can be entirely reprogrammed with the available tuning parameters. With the advent of frequency-agile communications and radar systems such as cognitive radios[1], there is an increasing need for extremely *programmable* signal filtering. While the primary considerations for mass-market applications (e.g., phased arrays, cell phones, base stations) relate to cost and footprint, these criteria are at most secondary considerations in other important applications which require high-performance filters with extreme tunability. For example, scientific exploration (e.g., space observation) and defense systems have strong needs for such agile free-form signal filtering.

Here, we depart from the established rational filter synthesis approach, and establish an alternative design paradigm that appears ideally suited to accommodate agile free-form filtering because it imposes virtually no *a priori* restrictions on the design space. Instead of selectively coupling spatially distinct resonators, we work with a quasi-2D metallic wave-chaotic cavity featuring many spatially overlapping resonances. This cavity backs an all-metallic programmable metasurface (a collection of meta-elements with individually reprogrammable scattering properties, details below) and is connected to multiple input/output ports. Reverberant scattering inside the low-loss chaotic cavity provides strong coupling between all ports and meta-elements, such that the configuration of the meta-elements efficiently controls the transfer function between the ports. Key advantages of the all-metallic metasurface include tunability across several GHz (covering the X band and the lower part of the Ku band), very low absorption, and the ability to sustain high signal powers. Our earlier works[2,3] used conceptually similar structures based on a 3D cavity and a printed-circuit-board programmable metasurface to demonstrate programmable signal differentiation and routing, but these works did not establish the connection to the broader scope of agile free-form signal filtering, for several reasons (discussed below) these experimental realizations were not suitable for most filtering applications, and the system's physics fundamentally differed since it operated in the regime of strong spectral modal overlap. In the present work, by optimizing the metasurface configuration in situ, we demonstrate agile ultra-wideband (UWB) high-fidelity control over reflectionless and transmissionless scattering modes. Along the way, we also experimentally verify recent theoretical predictions about their statistical properties and experimentally detect transmissionless exceptional points for the first time. We go on to implement the first demonstration of low-loss, programmable, reflectionless signal routing. Then, we investigate the tradeoff between routing fidelity and bandwidth. Finally, we demonstrate UWB tunable multi-band filters for which all features (band center, band width, reject or pass) can be reprogrammed.

**Conventional rational filter synthesis.** Design-space exploration and direct optimization purely based on computational electromagnetic solvers is computationally prohibitively costly. For this reason, as well as its compatibility with rational design principles, conventional filter synthesis leverages surrogate circuit-theory models that approximate constituent resonators as lumped (point-like) elements[4,5]. To roughly satisfy this approximation, irrespective of the targeted filter implementation (lumped or distributed), spatially disjoint resonators are considered. Note that this choice drastically limits the design space because it excludes by construction any structure with multiple spatially overlapping resonances, such as the system that we consider in this work. The spatially disjoint resonators are assumed to be uncoupled by default, and based on rational design approaches, desired couplings between selected pairs of resonators are added. A key conceptual tool in filter synthesis is the coupling matrix representation[6,7] (CMR) that captures the local resonant properties of all entities (internal resonators and ports), as well as their couplings. CMR bears significant conceptual links to the temporal coupled-mode theory[8] (TCMT) used in other communities (although these links have to date not been noticed to the best of our knowledge).

The rationally obtained design is usually fine-tuned during design closure to fix discrepancies between the actual and the targeted filter responses, arising from the assumptions underlying the utilized surrogate circuit-theory models. Among many design closure techniques (including recent efforts in the wave physics community[9]), the port tuning method[10–12] is particularly noteworthy in the context of the present paper. For this method, first, multiple auxiliary ports are defined within the optimized structure. Second, the structure is simulated in an electromagnetic solver, yielding a network description of the structure via an augmented scattering or impedance matrix, where the number of ports is the sum of the number of the structure's external and auxiliary internal ports. The structure itself is hence now treated as a black box that creates complex coupling mechanisms between, in principle, *all* ports. This is also a circuit-theory model of the structure, however, this time without relying on approximating constituent resonators as lumped elements. The effect of the constituent resonators appears only indirectly in the mutual coupling between ports. Based on the augmented scattering matrix, loads terminating the auxiliary ports can be optimized so that the scattering between the external ports approaches the targeted filter response. The optimized loads are chosen to correspond to practically realizable loads such as those of a microstrip of a given length or a lumped element.

While the above-summarized conventional rational filter synthesis has enabled the successful conception and prototyping of countless *static* filters, it appears very challenging to extend it to achieve filters with *significant reprogrammability*. Examples targeting limited reprogrammability such as center-frequency tuning of a fixed filter shape[13] or switching between a few filter shapes for the same center frequency[14] exist. But, in fact, even these limited types of reconfigurability encounter fundamental limitations. Center-frequency tuning of a single-frequency notch filter is arguably one of the simplest reconfigurability problems[15] but existing works only achieve tunability at the expense of filter shape fidelity[13]. A perfect single-frequency notch filter requires a zero of the filter transfer function on the real frequency axis at the desired frequency. For a static notch filter, despite challenges with fabrication inaccuracies and environmental perturbations, this can be achieved. By integrating some tunable components into the system, one can certainly move the zero around, but nothing constrains its motion to the real frequency axis,[^1] so it typically will move into the complex plane. This explains why simple tunable notch filters achieve tunability only at the expense of deteriorating shape fidelity.

---

[^1]: Theoretically, reflection [transmission] zeros of *PT*-symmetric [*T*-symmetric] structures are constrained to the real frequency axis or must exist in complex conjugate pairs off the real axis. (Sometimes, as parameters are tuned, these zeros meet on the real frequency axis at an exceptional point before becoming complex conjugate pairs)[16–19]. However, any realistic passive linear device has a finite amount of loss which inevitably breaks exact *PT* symmetry and *T* symmetry.

**Programmable metasurface with reverberation-induced non-locality.** Recently, perfect tunable notching has been achieved in Refs.[2,20] with a completely different approach (similar to our approach in the present work) in the context of coherent perfect absorption (CPA). CPA theory[21] describes very general conditions under which a lossy system can perfectly absorb an incident wavefront, including systems with multiple ports and/or spectrally overlapping resonances where a simple critical-coupling[22–24] picture fails. Perfectly incoming solutions ubiquitously exist in the complex frequency plane, corresponding to certain eigenfunctions. CPA occurs when such a solution is tuned to the real frequency axis and the injected wavefront (in the case of a multi-port system) is the appropriate eigenfunction. For such solutions the full scattering matrix has a zero eigenvalue at the targeted real frequency. To achieve CPA *at a desired frequency*, a massive parametrization of the system was necessary in practice[2,20]. This was experimentally achieved for the first time in Ref.[20] by placing a programmable metasurface inside a chaotic cavity; the scattering properties of each meta-atom were individually reconfigurable by controlling the bias voltage of a varactor. A similar approach was used in Ref.[2] to achieve tunable high-fidelity wave-based signal differentiation because the required transfer function is exactly that of a perfect single-frequency notch filter.

A conceptual understanding of the tunability of such complex scattering systems has only emerged within the last year in yet a different application context: smart radio environments. The latter, envisioned as pillar of next-generation wireless networks, endow the wireless channels between transmitters and receivers with tunability by placing a programmable metasurface on a wall in the environment[25–28]. Conceptually, the physics at play is the same as in the tunable CPA experiments mentioned above. The key question is how the scattering matrix defined by the antennas depends on the metasurface configuration. Given the overwhelming complexity of unknown geometrical and material details in the propagation environment, a brute-force discretization of the system for subsequent simulation with a computational electromagnetic solver is not possible. However, compact physics-compliant forward models can be formulated in terms of coupled dipoles[29] or, equivalently, circuit theory[30,31], and their parameters can be frugally estimated without detailed knowledge of the complex environment[29,31]. The circuit-theory version treats the radio environment as a black box and considers the antenna ports as well as auxiliary ports at the locations of the tunable lumped elements of each meta-atom that are to be terminated by tunable loads. The complex environment creates, in general, significant coupling effects between all ports (antenna ports and auxiliary lumped ports). Clearly, this approach is conceptually similar to the port tuning method during the design closure of filter synthesis (although this analogy has not been noticed previously). Importantly, formulating a forward model with CMR or TCMT appears quite difficult: any change of the metasurface configuration in principle alters all resonances and all couplings between all pairs of resonances and ports in the system in a seemingly unpredictable manner.[^2]

Although wave propagation in the system is linear, the dependence of the scattering matrix on the metasurface configuration is non-linear because of multi-bounce paths that encounter multiple meta-atoms; these multi-bounce paths are compactly captured by a matrix inversion in physics-compliant models[33]. The stronger the reverberation inside the environment is, the more significant multi-bounce paths exist. In other words, there are substantial reverberation-induced non-local interactions between the programmable meta-atoms which boost the control of the

---

[^2] Within the context of *static* diffractive nonlocal metasurfaces (distinct from the *chaotic-cavity-backed* and *programmable* metasurface considered in the present work), an extension of TCMT to account for spatially dispersive resonances via spatially varying modal parameters has recently been explored in Ref.[32].

metasurface configuration over the system's transfer function.³ There is no unique definition of "non-locality" in the literature, but the general idea is that spatially separated entities significantly interact, such that the "local" response at a given location depends significantly on the fields and/or structures at distant locations and is hence "non-local". Distinct from other works on non-local metamaterials and metasurfaces[19,36–40], in our chaotic-cavity backed programmable metasurface the non-locality originates chiefly from reverberation.

It is equally valid to describe our system as a "programmable chaotic cavity". However, we prefer the term "chaotic-cavity-backed reverberation-non-local programmable metasurface" for two distinct reasons. On the one hand, the study of shape-tunable cavities is not always associated with optimization to achieve a desired scattering functionality. Indeed, for decades shape-tunable cavities have been predominantly explored in terms of their *statistical* properties in diverse contexts ranging from mesoscopic systems[41,42] to electromagnetic compatibility[43]. The use of the tunability for optimization is more recent[44]. On the other hand, physical models that can be calibrated to deterministically predict the scattering properties of a given cavity configuration view the ports and meta-elements as the primary entities[29–31], and the presence of the cavity appears in the coupling of these primary entities (see discussion below). Hence, in order to emphasize the use of the programmability for optimization, and in light of how physical models of the system can be formulated, we refer to our system as "chaotic-cavity-backed reverberation-non-local programmable metasurface".

**Agile free-form signal filtering.** The apparent strength of the approach based on a chaotic-cavity-backed programmable metasurface is that, in principle, the design space is completely unconstrained. The underlying resonances can be arbitrarily spatially extended, and every tunable degree of freedom can, in principle, simultaneously affect all resonances and all coupling coefficients. For this reason, the chaotic-cavity-backed programmable metasurface appears very promising for agile free-form signal filtering. By "free form", we mean that in principle (almost) any filter response (within physical bounds) can be realized. By "agile", we mean the possibility of arbitrarily reconfiguring the filter response in situ. However, the feasibility of this vision remains to date unexplored. So far, Refs.[2,20] only demonstrated perfect tunable single-frequency notching in reflection, and Ref.[3] extended the concept to single-frequency programmable reflectionless signal routing (we discuss the underlying theory in Sec. 4). As noted above, in Ref.[3], switching between a few single-frequency routing functionalities was demonstrated, but desired transmissions suffered from ~20 dB attenuation. Moreover, all of these studies on scattering singularities like CPA were limited to single-frequency filtering whereas realistic signals often have a finite bandwidth. Furthermore, the utilized PIN-diode-programmable (or varactor-diode-programmable) resonant meta-atoms have several inevitable drawbacks: they only significantly manipulate electromagnetic waves within an interval of a few hundred MHz, they inevitably attenuate waves (although Ohmic losses on the cavity walls were more significant in Refs.[2,3,20]), the PIN diodes (or varactors) consume power to maintain a given configuration, and the PIN diodes (or varactors) cannot handle strong power without risking to generate their own harmonics or intermodulation distortions. Finally, the 3D setups in Refs.[2,3,20] were quite bulky.

Here, we overcome all of the above-listed technological limitations with a mechanically programmable all-metallic metasurface backed by a quasi-2D chaotic cavity. Our device offers UWB tunability of several GHz, only minimally attenuates desired transmissions (typically by roughly 1 dB), can handle the levels of incident signal power typically encountered in defense applications, only consumes power to reconfigure the filter function, is significantly more compact, and is compatible with conformal design constraints. The significantly longer reverberation times (due to the drastically lower level of absorption) imply that fewer tunable degrees of freedom are needed for a desired level of wave control because the field is more sensitive; in addition, in the present work the degrees of freedom are continuously tunable rather than 1-bit programmable, offering access to additional independent states. These major technological improvements enable the first demonstrations of UWB programmable control over transfer function zeros and low-loss programmable reflectionless signal routing. Moreover, for the first time, we go beyond the single-frequency regime by studying the trade-off between routing functionality and bandwidth, as well as to what extent multi-band filters with arbitrarily located pass and stop bands can be implemented.

---

³ The connection between reverberation (dwell time) and field sensitivity to perturbations was established in other contexts, for instance, in Refs.[34,35].

## 2. Implementation and In Situ Optimization of the Prototype

To start, we introduce and characterize our all-metallic reverberation-non-local programmable metasurface. Our device shown in Fig. 1a is built upon a D-shaped[45–48] chaotic cavity which is quasi-2D because of its sub-wavelength vertical dimension. Three coax-to-waveguide transitions are connected at arbitrarily chosen locations to the cavity, constituting three ports. Each programmable meta-element is a cylindrical metallic shaft with threaded grooves that can be sunk into the cavity at a continuously controllable depth from the upper surface using a stepper motor. The locations of the $N = 14$ meta-elements are arbitrarily chosen. Although a few mechanically reconfigurable metasurfaces exist in the literature[49–51], these devices were conceived for reflect-array applications and drastically differ from ours; most importantly, these devices did not feature any reverberation non-locality, nor were they used for signal filtering. Within the filter literature, motorized mechanical reconfigurability appears to be uncommon but a few examples of mechanically reconfigurable filters (within the conventional rational synthesis paradigm and not free-form reconfigurable) based on manual control[52], liquid metal[53], or MEMS[54] exist.

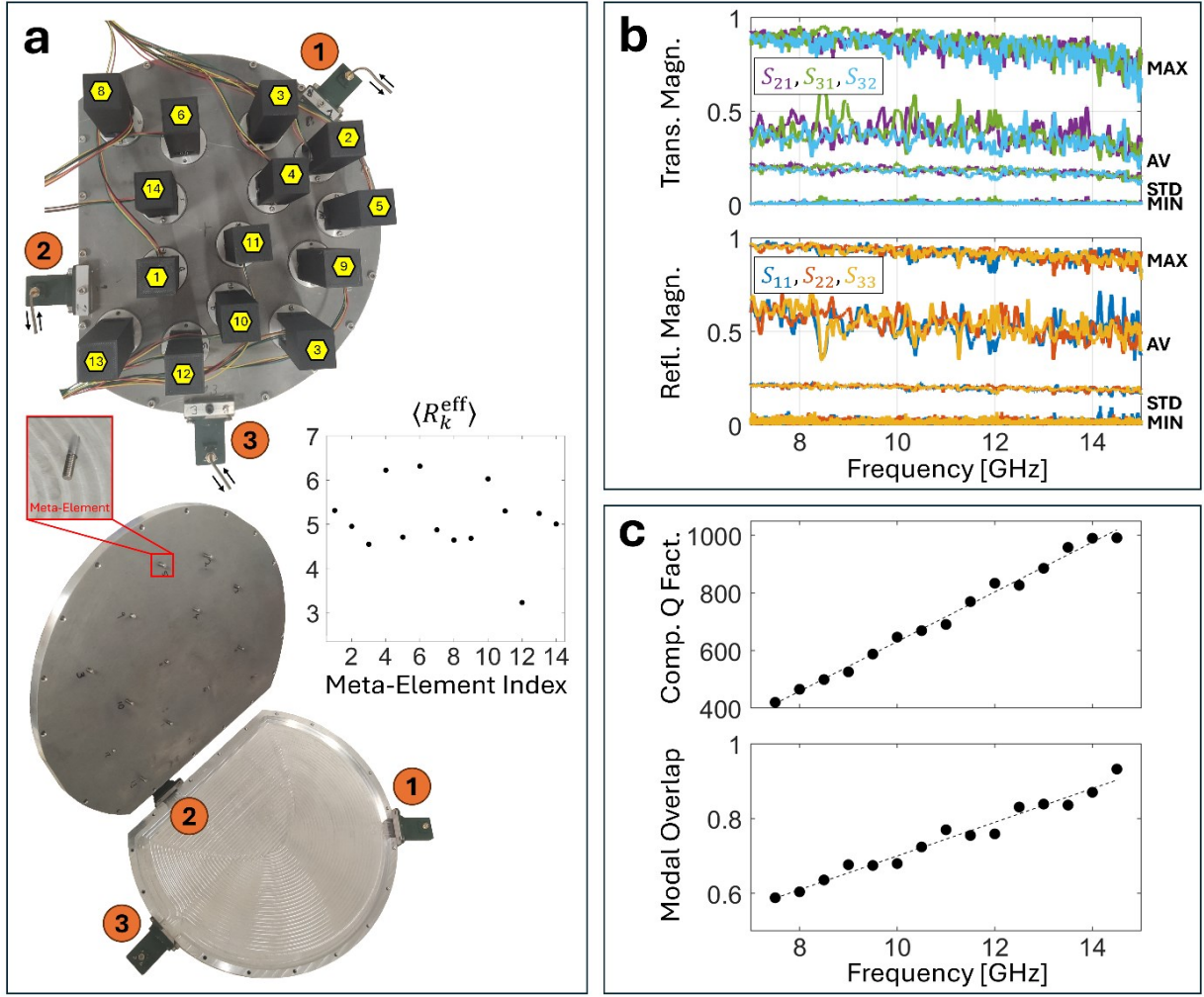

**Figure 1. Implementation and characteristics of quasi-2D chaotic-cavity-backed all-metallic programmable metasurface with three input/output ports.** (a) Photographs of the experimental prototype (top view and inside view). The inset shows one of the meta-elements. The estimated number of "independent" states (quantified by $\langle R_k^{\text{eff}} \rangle$, see Methods) of each of the 14 meta-elements is also shown. (b) Based on 2040 random metasurface configurations, the values of $\max_{\text{rea}}(|S_{ij}|)$, $|\langle S_{ij} \rangle_{\text{rea}}|$, $\text{SD}_{\text{rea}}(|S_{ij}|)$, and $\min_{\text{rea}}(|S_{ij}|)$ are shown (denoted by MAX, AV, STD, MIN) for all transmission and reflection coefficients. (c) The estimated composite quality factor and modal overlap are plotted as a function of frequency (evaluated across 1 GHz intervals, see Methods).

As seen in Fig. 1b, the configuration of the programmable metasurface offers a strong control over the transfer function within an UWB range. The smallest magnitude of each scattering coefficient seen across 2040 random metasurface configurations approaches zero at all frequencies between 7.5 and 14. GHz, while the largest magnitude is close to unity (decays from roughly 0.9 [-1 dB] to 0.7 [-3 dB] in transmission and from roughly 0.96 [-0.4 dB] to 0.87 [-1.2 dB] in reflection between 7.5 and 14. GHz). Note that optimized configurations can go beyond these values. Importantly, signal attenuation due to Ohmic losses inside the system is fairly weak (in contrast to Refs.[2,3,20]) and only slowly increases with frequency. The system's composite quality factor[55] scales roughly linearly with frequency from around 420 at 7.5 GHz to around 1020 at 14.5 GHz, as seen in Fig. 1c. Meanwhile, the amount of absorption increases with frequency (as expected), as seen by the frequency dependence of the maximum transmission values across random configurations in Fig. 1b and the fact that the modal overlap (see Methods for estimation procedure) slightly increases from 0.6 to 0.9 in the interval from 7.5 GHz to 14.5 GHz, as seen in Fig. 1c. Throughout this work, we are hence operating close to the regime of isolated resonances – in contrast to Refs.[2,3,20] which featured very large modal overlap due to the much stronger absorption. The number of "independent" states to which a given continuously tunable meta-element can be configured is around 5 on average, as seen in Fig. 1a (see Methods for estimation procedure).

Our goal in the present paper is to experimentally demonstrate that our proposed hardware enables agile free-form signal filtering. We began with rather straightforward optimization methods and found them to be successful. Specifically, we use closed-loop iterative optimizations relying on in situ experimental measurements: every time a forward pass (determining the scattering matrix for a given configuration) is needed, we directly measure the sought-after information experimentally. Details of our optimization algorithm are provided in the Methods. Although least prone to modelling and parameter-estimation errors, this approach works in closed form and is inevitably slow. Future work will tackle the parameter estimation of a physical model of our system to unlock open-loop control: given a physical model of the system, an optimized system configuration to achieve any desired functionality can be identified in software without requiring further access to the experiment. We believe that such open-loop control can be accomplished by generalizing the physical model formulated in Ref.[29] to describe the meta-elements of our present system via a multi-polar representation[56,57] (truncated at finite order), or, equivalently, as a collection of $m$ non-local dipoles[58]. Thereby, the model will feature two types of non-locality: within each meta-element (this is the usual type of non-locality[40,58,59]) and between all entities due to reverberation (this is the unconventional type of non-locality).

## 3. UWB Control over Transfer Function Zeros

Having introduced our reverberation-non-local programmable metasurface prototype, we now probe its control over the system's transfer function, i.e., its suitability for free-form signal filtering. In this section, we focus on control over the transfer function zeros. Within the realm of microwave network theory which applies to our system with discrete point-like ports, it is customary to factorize the determinant of a system's continuous-time transfer function (the entire scattering matrix or a square block thereof) into the ratio of two polynomials of the complex frequency variable[60,61], identifying poles and zeros in the complex-frequency plane at which the determinant tends to infinity or zero, respectively. The thereby identified set of poles and zeros compactly captures important information about the system's scattering properties[62,63].[4] Our focus on transfer function zeros is doubly motivated. On the one hand, recently their technological relevance has clearly emerged, e.g., for signal differentiation[2] and for achieving full phase modulation while maintaining close-to-unity amplitude in free-space optical metasurfaces[66–68]. On the other hand, there exist recent rigorous scattering theory approaches which frame transfer function zeros as eigenvalue problems related to an underlying wave operator with reflectionless scattering boundary conditions (i.e., for zeros of the determinant of diagonal blocks of the scattering matrix[16]) and also for scattering boundary conditions which impose zeros of a transmission matrix (i.e., a square off-diagonal block of the scattering matrix[17]). In both cases, typically, many solutions exist at discrete frequencies in the complex frequency plane, but they only correspond to zeros in steady-state scattering when they occur or are parametrically tuned to a real frequency. Our experimental results on the control of transfer function zeros will be interpreted in light of these related theoretical predictions. Both reflection and transmission zeros can be understood as special cases of a more general concept of coherent perfect extinction (CPE)[69].

---

[4] In optical scattering problems involving free-space excitations, more general formulations are often required for an accurate description[62,64,65].

We first analyze unconstrained reflection and transmission zeros by determining the rate at which they occur, anywhere on the real frequency axis (hence "unconstrained"). The occurrence rates are estimated based on scattering measurements for 2040 random metasurface configurations and 13618 frequency points between 7 and 15 GHz. Because of experimental limitations, the imaginary part of the frequency at which a zero exists will never be exactly zero so that in practice we determine the rate at which (reflected or transmitted) output powers fall below a certain threshold. Obviously, the occurrence rate is lower for more stringent thresholds; based on Fig. 2a, it appears that the scaling of the occurrence rate with the threshold value is generically exponential.

An $r$-port reflectionless scattering mode[16] (RSM) occurs when a diagonal $r \times r$ block[5] of the scattering matrix (denoted by $\mathcal{R}$) has a zero eigenvalue; its excitation requires imposing the associated eigenvector as the input wavefront, implying that the system is only excited via the $r$ selected ports. Since our system has $n = 3$ ports, we can study 1-port RSMs ($r = 1$) and 2-port RSMs ($r = 2$), as well as CPA ($r = n = 3$). RSMs can occur in lossless systems for $r < n$, but CPA (a special case of RSM with $r = n$) can only occur if the system features a finite amount of irreversible loss (conversion of wave energy to heat, etc.)[21]. In practice, we consider a coherently minimized reflected power[6] (i.e., the smallest eigenvalue of $\mathcal{R}^\dagger \mathcal{R}$) below the threshold as revealing an RSM. We observe in Fig. 2a that the occurrence rate of 1-port RSMs and 2-port RSMs is almost identical, and significantly higher than the occurrence rate of CPAs. These observations are in very good agreement with expectations from the RSM theory for the lossless case. If the scattering system has negligible loss, then every 1-port RSM has a 2-port counterpart at the same frequency obtained from time-reversal symmetry by simply interchanging the input and output ports (and imposing the correct coherent input for the two-port case). Hence these two rates should be identical in the lossless case, as we find (approximately) in our low-loss system. In addition, CPA requires irreversible loss processes such that it would be impossible in a lossless system; hence it is significantly rarer than 1-port RSMs and 2-port RSMs in our low-loss system.

By fitting the continuous-time frequency-dependent scattering data, as mentioned, we can also study the zeros when they are off the real frequency axis for a selected realization and frequency interval (as in Fig. 2b) as well as their statistics over many realizations and frequency intervals (as in Fig. 2c). Statistically, the larger $r$ is, the higher up in the complex plane we expect the zeros of $\mathcal{R}$ to lie. Neglecting the small absorption and assuming that the scattering loss at each port is roughly the same, we expect the 1-port case to be overdamped, with a zero in the lower half plane and the corresponding 2-port case to be underdamped, with a zero at the complex conjugate frequency in the upper half plane by time reversal. In our system, attenuation is sufficiently small to qualitatively confirm this expectation upon visual inspection of Fig. 2b. The empirically determined statistical distribution of the imaginary parts of the zeros shown in Fig. 2c corroborates this fact. The two distributions are roughly symmetrical with respect to a value of the imaginary part slightly below zero because of the small amount of absorption loss which shifts both distributions slightly toward negative values. This also explains why the occurrence rate of 2-port RSMs very slightly exceeds that of 1-port RSMs in Fig. 2a. In contrast to 1p-RSMs and 2p-RSMs, CPA can only occur if the absorption is sufficient to pull a full S-matrix zero down to the real frequency axis at some frequency and for some input wavefront. This is possible in our system, but significantly less likely (see Fig. 2a), because of the low absorption loss. As seen in Fig. 2c, the distribution of the imaginary part of the zeros of the full scattering matrix is significantly skewed toward positive values, and only a small tail reaches below zero. This is in line with our observation from Fig. 2a that CPA is significantly less likely than the other RSMs. (As noted, in the limit of zero absorption, CPA would be impossible.) Interestingly, in Fig. 2b ("CPA: {1,2,3}"), the poles and zeros of the complete scattering matrix are seen to be roughly complex conjugates, which would be exactly true in the absence of loss.

---

[5] The indexing of the system ports is arbitrary, such that a choice of $r$ non-contiguous port indices can also be represented as selecting a diagonal block of the scattering matrix upon altering the port index order.

[6] See also related discussion in Ref.[3] about the generalization of the coherently enhanced absorption from Ref.[70] based on the entire scattering matrix to the coherently minimized reflection based on a diagonal block of the scattering matrix.

The present work constitutes the first experimental analysis of these unconstrained RSMs and their statistics in a tunable *low-loss* system, and the first confirmation of the behavior expected from the RSM/CPA theory. The observations in Fig. 2a differ qualitatively from the results of a similar analysis in Ref.[3] for a highly overdamped tunable system. In Ref.[3], the occurrence rate of an $r$-RSM increased with the value of $r$, CPA hence being the *most likely*. Recall that, statistically, the larger $r$ is, the higher in the complex plane the zeros of $R$ lie. In Ref.[3], the strong absorption damping moved the distributions of the $\mathcal{R}$-zeros well below the real frequency axis, such that distributions corresponding to larger values of $r$ were closer to the real frequency axis, and hence CPA was the most likely type of RSM in Ref.[3].

We also observe in Fig. 2a that the occurrence rate of two *simultaneous* 1-port RSMs (i.e. two 1-port RSMs with different choices of input ports at the same frequency) is similar to that of two *simultaneous* 2-port RSMs, and much lower than for single RSMs. The approximate equality of these two rates is again expected from the approximate time-reversal symmetry of our low-loss system (small differences in the two rates can be attributed to uncertainties due to the limited sample size). We furthermore checked for three simultaneous RSMs at the same frequency and degenerate RSMs but did not find any in our data. Finally, for lenient threshold values, we found some simultaneous 1-port RSMs and matching TSMs (i.e., input via the port involved in the 1-port RSM) at the same frequency, which constitute instances of CPE and correspond to transfer functions relevant to the signal routing studied further below.

We now turn our attention to transmissionless scattering modes (TSMs) which fundamentally differ from RSMs in important ways. TSMs are $\mathcal{T}$-zeros (i.e., zeros of the determinant of an off-diagonal square block of the scattering matrix) located on the real frequency axis[17]; in our 3-port system, we are limited to considering zeros of individual transmission coefficients. Unconstrained TSMs appear to occur more frequently than unconstrained RSMs according to Fig. 2a. This is in line with theoretical expectations. Indeed, under exact *T*-symmetry (i.e., for a lossless 2-port system), $\mathcal{T}$-zeros would in fact be constrained to existing as TSMs on the real-frequency axis or as complex conjugate pairs[17]; similar constraints apply to $\mathcal{R}$-zeros only under the more stringent *PT*-symmetry constraint[16,18,19]. We observe indeed in Fig. 2c that the distribution of imaginary parts of the $\mathcal{T}$-zeros extracted from our experimental data is sharply peaked just below zero, due to the small amount of absorption. It is further predicted by theory that by continuously tuning one parameter of a lossless system, one should be able to observe the coalescence of two TSMs on the real-frequency axis, constituting a transmissionless exceptional point (EP), before they become a complex-conjugate $\mathcal{T}$-zero pair[17]. Although the weak absorption in our system breaks exact *T*-symmetry, we are able to detect the expected $\mathcal{T}$-zero motion, displaced slightly into the lower half complex frequency plane as seen in Fig. 2d. Specifically, as the perturbation is gradually turned on by sinking a meta-element into the cavity (all other meta-elements are fixed in a random configuration and the third port is terminated with a perfectly reflecting open-circuit load), two almost complex conjugate $\mathcal{T}$-zeros near 10.59 GHz approach the real frequency axis and anti-cross just below it (they would meet on the real frequency axis and slide along it in opposite directions in the absence of absorption). Because exact *T*-symmetry is broken, two-parameter tuning would be required for them to exactly meet. Then the $\mathcal{T}$-zero which moves to higher frequencies at higher perturbation strength undergoes a second similar anti-crossing with a different $\mathcal{T}$-zero partner at around 10.61 GHz. This scenario shows the interesting richness of our tunable chaotic cavity, as the first near-EP corresponds to the restoration of (approximate) *T*-symmetry, and the second one to the breaking of it. Analogous motion of $\mathcal{R}$-zeros under (approximate) *PT*-symmetry was experimentally detected in Ref.[19] (see Fig. 3 therein). The present results from Fig. 2d are the first approximate experimental observation of transmissionless EPs in general, and in a chaotic system in particular.

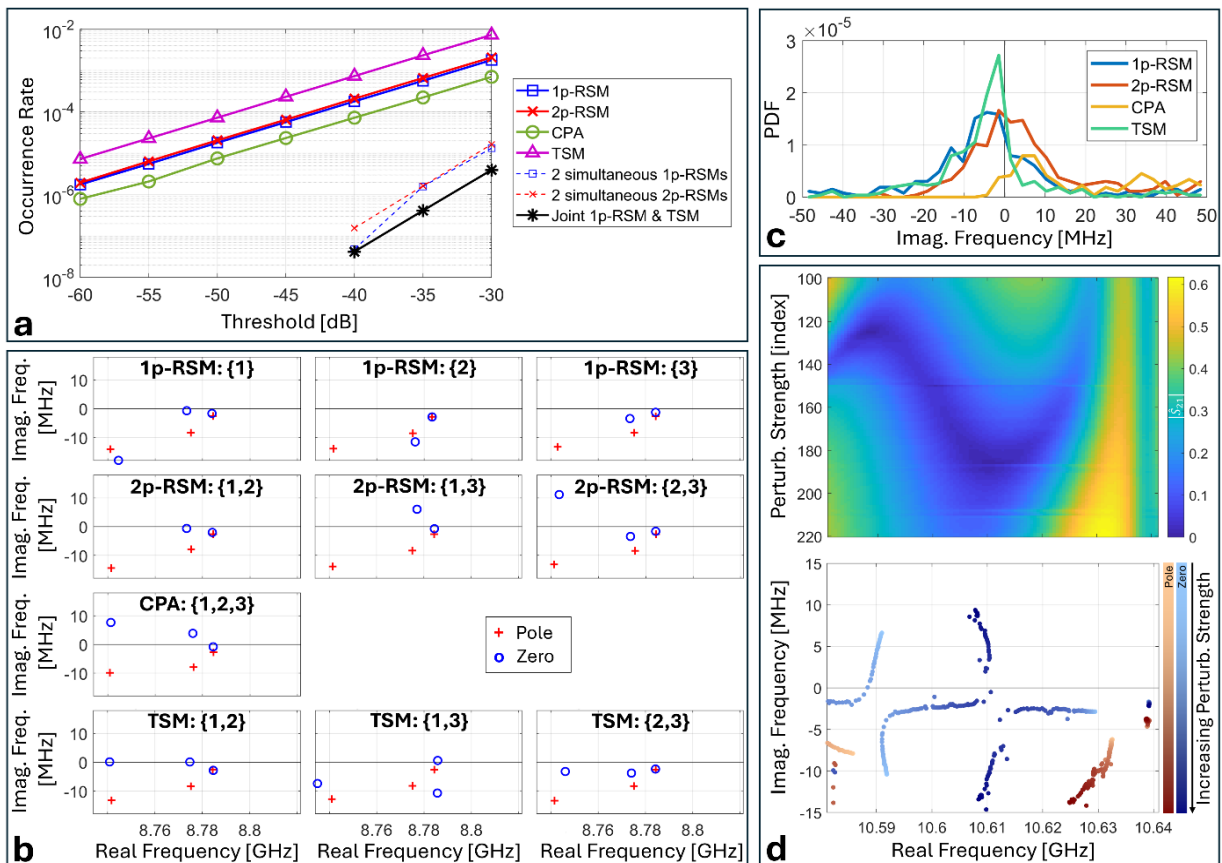

**Figure 2. Unconstrained transfer function zeros.** (a) Occurrence rate of transfer function zeros (as a function of the threshold value for scattering suppression), evaluated across 2040 random metasurface configurations and 13618 frequency points between 7 and 15 GHz. Besides 1-port RSMs, 2-port RSMs, CPAs and TSMs, we also plot the occurrence rate for two simultaneous 1-port RSMs at the same frequency and two simultaneous two-port RSMs at the same frequency, as well as the occurrence rate for a joint 1-port RSM and a matching TSM (i.e., input via the port featuring the 1-port RSM) at the same frequency. We furthermore searched for three simultaneous RSMs and degenerate RSMs but did not find any instances in the data set. (b) Example of complex plane frequency eigenvalues (poles and zeros) obtained by fitting scattering data for different reflection and transmission spectra within a narrow frequency range for a random metasurface configuration. The indices in curly brackets indicate the involved ports. (c) Probability density function (PDF) of the imaginary part of different transfer function zeros, evaluated based on the scattering matrix measured for ten random metasurface configurations between 7 and 15 GHz. (d) Selected example of a perturbation-frequency map of a transmission coefficient amplitude when the third port is terminated by an open-circuit load. A random metasurface configuration is chosen, and then one meta-element is gradually displaced into the cavity, while measuring the resulting scattering matrix for every step.

The corresponding motion of poles and zeros in the complex plane is also indicated. The shade of the dots encodes the depth to which the continuously tuned meta-element has been sunk into the cavity. The anti-crossings of transmission zeros near 10.59 GHz and 10.61 GHz correspond to approximate transmissionless exceptional points.

So far, we have only considered unconstrained transfer function zeros occurring anywhere along the real frequency axis. Since practical systems are usually intended to operate at specified frequencies, we now turn our attention to frequency-constrained transfer function zeros that lie at a desired location on the real frequency axis. As seen in Fig. 2a, the probability that a random metasurface configuration yields the desired frequency-constrained RSM or TSM is vanishingly small (e.g., around $10^{-5}$ for 1p-RSMs and 2p-RSMs if we fix a threshold of -50 dB). However, by optimizing the metasurface configuration, we can impose these zeros at desired locations on the real frequency axis. We display in Fig. 3 a few representative examples demonstrating that the control offered by our metasurface is sufficient to impose 1-port RSMs, 2-port RSMs, CPAs or TSMs at any desired frequency within the UWB range of operation of our device, achieving undesired output powers well below -50 dB. While Ref.[2] did achieve frequency-constrained 1-port RSMs at desired frequencies within a 400 MHz interval, here we achieve frequency-constrained 1-port RSMs and other interesting transfer function zeros over a much wider 7 GHz interval. Thus, following Ref.[2], and based on the results displayed in Fig. 3, our system could be directly used as an UWB meta-programmable signal differentiator, capable of operating with high fidelity in reflection (with single or multi-channel coherent input) or transmission. We also optimized the metasurface configuration to yield two simultaneous one-port RSMs at the same desired frequency; the probability of occurrence thereof is yet orders of magnitude lower according to Fig. 2a. While we do manage to significantly suppress two single-port reflection coefficients at the same desired frequency simultaneously, this objective is apparently harder than the single RSMs or TSMs, and the reflected power does not drop below -50 dB in the case of two simultaneous 1-port RSMs.

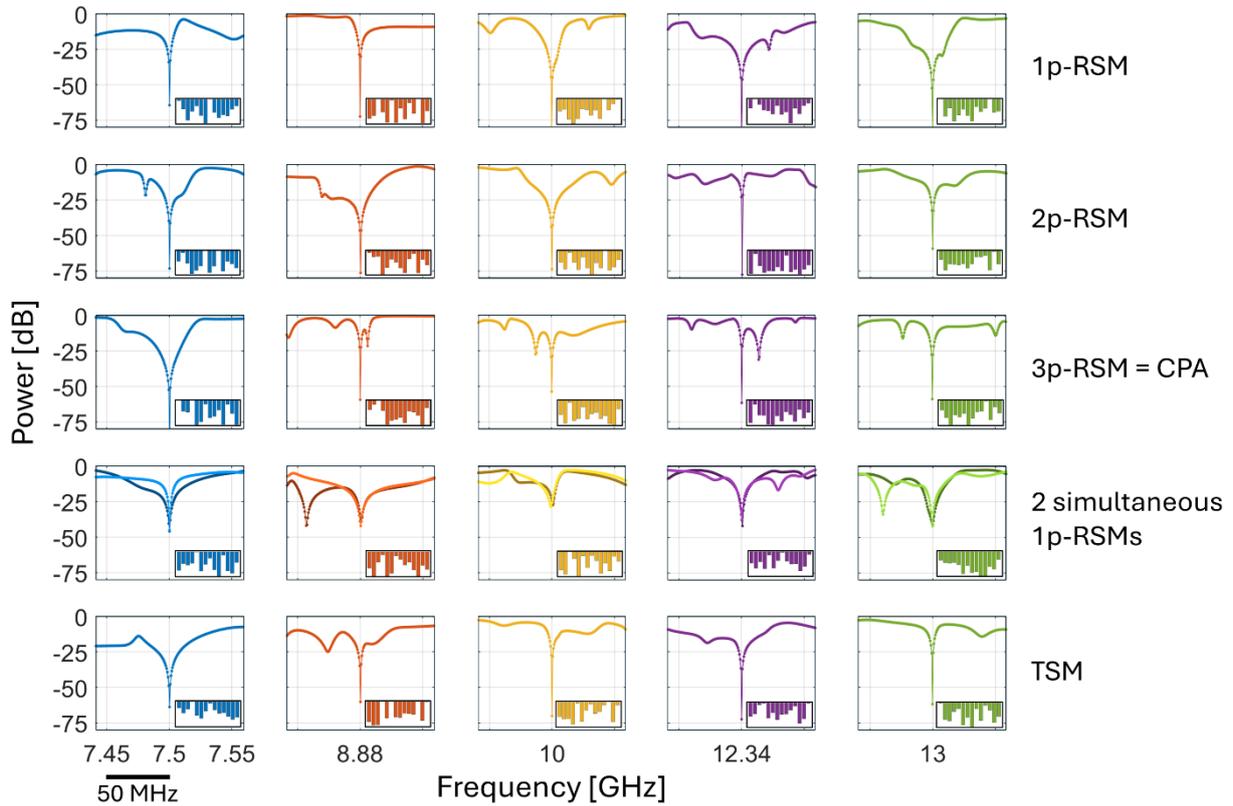

**Figure 3. Frequency-constrained programmable transfer function zeros.** Demonstration of UWB control over transfer function zeros by imposing RSMs and TSMs at desired frequencies (shown here for five arbitrarily chosen frequencies). The vertical axis displays the reflected or transmitted power. The insets display the corresponding metasurface configuration (meta-elements sorted by index). The selected involved ports are {1}, {1,2}, {1,2,3}, {1}, {1,2}. Similar results obtained for other choices of involved ports and target frequencies are not shown for conciseness.

## 4. Low-Loss Reflectionless Signal Routing with Functionalized CPEs

Having investigated unconstrained and frequency-constrained RSMs and TSMs in the previous section, we now consider a "functionalized" combination of the two for signal routing. In order to reflectionlessly route a signal at a desired frequency in our 3-port system from a desired input port to a desired output port, three constraints must be satisfied simultaneously: (i) a 1-port RSM at the input port, (ii) a TSM from the input to the undesired output port, and (ii) maximal transmission from the input to the desired output port. Whereas the first two constraints relating to minimizing output power can be framed as a coherent extinction problem, the third constraint regarding maximization prevents us in general from being able to frame reflectionless signal routing as an eigenvalue problem, such that current scattering theory does not offer theoretical predictions in general for the "*functionalized* CPE" underlying signal routing.[7] Only if our 3-port system were lossless, the first two constraints combined with flux conservation would guarantee unit transmission into the desired port (i.e., the third constraint would be automatically satisfied if the first two constraints were satisfied). The probability that a random metasurface configuration simultaneously satisfies the first two constraints at a desired frequency in our system was seen to be extremely low in Fig. 2a (e.g., below $10^{-7}$ when we fix a lenient threshold of only -40 dB). However, once the first two conditions are satisfied, we expect it to be easy to additionally satisfy the third condition because our system is relatively low-loss. Selected examples in Fig. 4a illustrate that the available tunable degrees of freedom in our system are sufficient to simultaneously satisfy all three requirements, irrespective of the choice of input and desired output ports and irrespective of the chosen operation frequency. In all cases, we suppress undesired outputs (reflection at input port and transmission to undesired output port) by at least 39 dB; meanwhile, the desired transmission is attenuated by at most 1 dB, yielding a discrimination between desired and undesired outputs of at least 38 dB. Besides the UWB tunability of the operating frequency, we would here like to highlight the remarkably low attenuation of the desired transmission of only 1 dB (in contrast to roughly 20 dB in Ref.[3]). Thus, the routing performance achieved with our reverberation-non-local programmable metasurface, along with its almost arbitrary reprogrammability, shows that this approach has significant technological promise.

---

[7] As a side remark on terminology and context, we note that the functionalization of RSMs [which naturally satisfy constraint (i)] for signal routing by imposing additionally the constraints (ii) and (iii) was first studied in Ref.[3]. In light of the terminology of CPE later proposed in Ref.[69], signal routing can also be framed as functionalizing a CPE [which can naturally satisfy constraints (i) and (ii)] by additionally imposing constraint (iii). However, Ref.[69] did not study any functionalized CPEs.

To gain further insights, we numerically simulated a lossless 2D cavity that resembles our experimental prototype. In our simulations, the transfer function of the cavity is controlled by the orientation of ten elliptical scattering elements. A configuration of the simulated system optimized for reflectionless routing at a desired frequency $f_0$ is displayed in Fig. 4b together with the associated field map and scattering spectra. The achieved performance in terms of reflection suppression of 45 dB is comparable to our experimental results whereas the suppression of the undesired transmission is 41 dB. Meanwhile, the desired transmission closely approaches unity due to the lossless nature of the simulated system. We also optimized the simulated system toward a reflectionless demultiplexing functionality (simultaneous complementary routing at two desired frequency $f_1$ and $f_2$).[3] The obtained results, shown in Fig. 4c, suppress reflection and undesired transmission by at least 46 dB and 43 dB, respectively. Again, the desired transmission closely approaches unity.

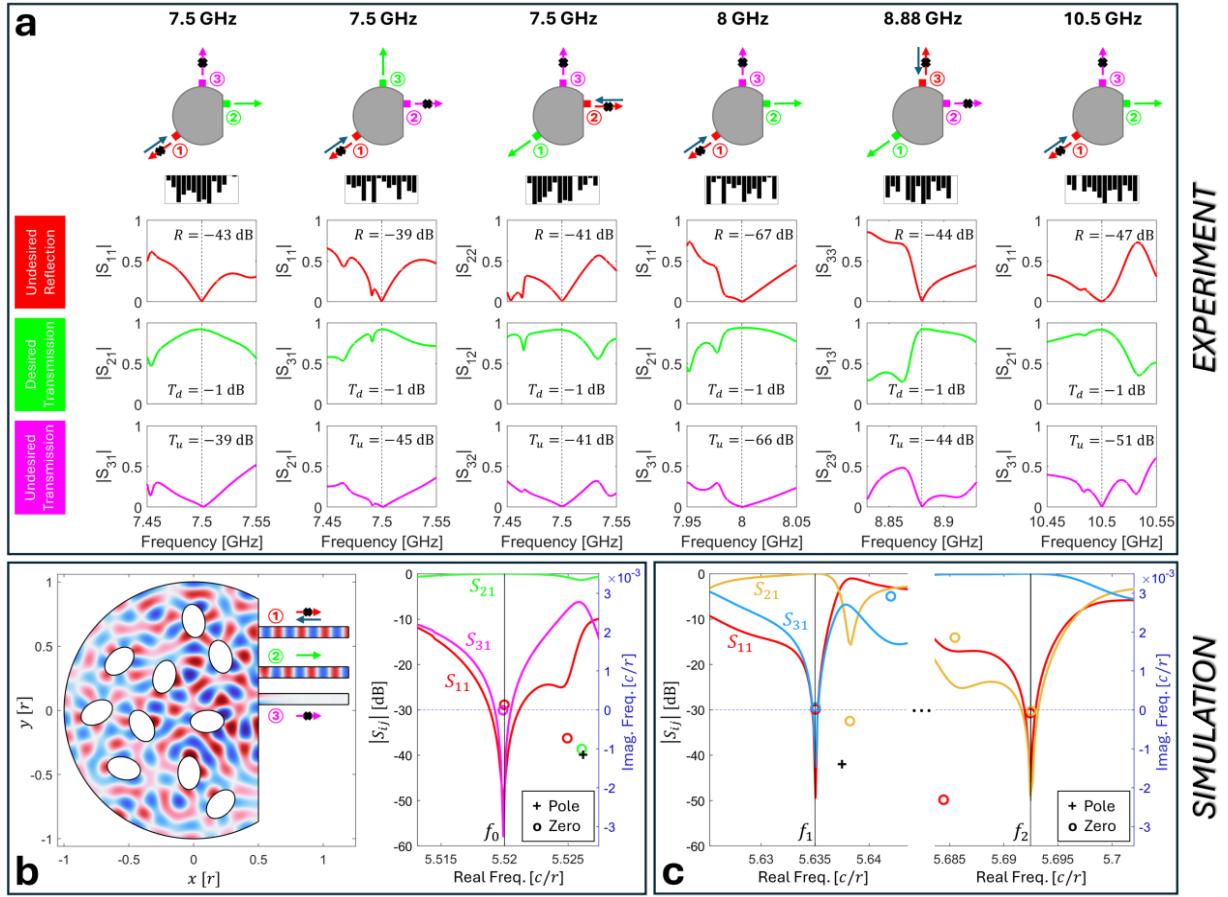

**Figure 4. Low-loss reconfigurable reflectionless signal routing.** (a) The same non-local programmable metasurface is reconfigured to implement six representative reflectionless routing functionalities, differing in the operating frequency and/or the input and (un)desired output ports. In each case, the reflection of the input port is suppressed by at least 39 dB, the undesired transmission is suppressed by at least 39 dB and the desired transmission is not attenuated by more than 1 dB. Hence, the discrimination is better than 38 dB in all cases. The insets on top display the desired routing setup and the corresponding optimized metasurface configuration (meta-elements sorted by index). Extracted scattering singularities for the example at 10.5 GHz are shown in Fig. 5a. (b) Numerical simulation of a lossless system similar to our experimental prototype, optimized (by tuning the orientation of ten elliptical scattering elements, see Methods) for reflectionless routing at $f_0$ from port 1 to port 2. The field pattern (real part of the out-of-plane component of the H-field), scattering spectra and extracted singularities are shown. Reflection and undesired transmission are suppressed by 45 dB and 41 dB, respectively. $r$ denotes the radius of the cavity, $c$ denotes the speed of light. (c) Numerical simulation of a configuration of the system from (b) optimized toward reflectionless demultiplexing ($f_1$ and $f_2$ are injected via Port 1 and supposed to only exit via Port 3 or Port 2, respectively). Reflection and undesired transmission are suppressed by at least 46 dB and 43 dB, respectively. Scattering spectra and extracted singularities are shown.

## 5. Trade-off between Routing Fidelity and Bandwidth

The reflectionless reconfigurable signal routers implemented in the previous section were designed for optimized operation at a single frequency, and only work well over a narrow bandwidth. However, many realistic signals have a substantial bandwidth, raising the question of to what extent (near-)reflectionless reconfigurable signal routing is possible for signals with finite bandwidth. The main obstacle to increased bandwidth relates to the desired signal extinctions: The various types of transfer function zeros discussed above (CPA, RSM, TSM, CPE) are all fundamentally resonant phenomena and hence only occur at discrete frequencies; the associated extinction dips are generally very narrow, as seen in Fig. 3. Theoretically, exceptional points (EPs) of such zeros[71,16,72,17–19] have a parametrically flatter frequency response such that high extinction (but not rigorously perfect extinction) would extend over a somewhat larger bandwidth. However, creating a single EP $\mathcal{R}$-zero on the real frequency axis requires tuning of three (sufficiently impactful) parameters and is not straightforward in a chaotic cavity such as ours. The difficulties of doing this simultaneously with other possibly conflicting constraints is not understood theoretically, so increasing bandwidth will have to be explored through empirical optimization. Such optimization will not likely lead to exact EPs but rather bring several of the requisite zeros into the desired bandwidth and close to, but not onto the real frequency axis. Conversely, poles of the scattering coefficients, which cause all scattering matrix elements to diverge, should be pushed out of the desired bandwidth. The goal to route finite-bandwidth signals seems to inevitably imply a trade-off between routing fidelity and bandwidth, which we now explore.

We systematically studied the achievable routing fidelity as a function of desired signal bandwidth for various center frequencies. Selected results for a center frequency of 10.5 GHz are shown in Fig. 5a. It is apparent that maximizing the desired transmission is indeed not a bottleneck. Up to roughly 20 MHz of bandwidth, the optimized desired transmission never drops below -1 dB, and even for larger bandwidths, it only deteriorates slightly as the bandwidth is increased. To gain further insights, we also extracted the singularities (poles and zeros) of each displayed scattering coefficient spectrum. For the optimized desired transmissions, we observe that all singularities are pushed out of the band of interest within the vicinity of the real frequency axis, enabling an almost flat desired transmission spectrum within the band of interest. In contrast, the minimum suppression of reflection and undesired transmission clearly deteriorates with operation bandwidth. For low desired bandwidths, a reflection zero and an undesired-transmission zero lie very near the central frequency almost on the real frequency axis, just as we expect for a deep dip. For moderate and larger bandwidths, as we anticipated, multiple such zeros lie in the vicinity of the real frequency axis to suppress the undesired outputs as much as possible, but far from perfectly. Meanwhile, since the poles are the same for all scattering coefficients, the reflection and undesired transmission spectra also do not feature any pole close to the real frequency axis within the desired band. These observations highlight that the real frequencies at which poles and various types of zeros occur can be drastically different, as recently emphasized in RSM theory[16]. We conjecture that the reverberation-induced non-locality engendered by our chaotic cavity is what allows us to tailor the zeros and pole locations of the scattering coefficients in this highly non-trivial manner.

In Fig. 5b, we study the tradeoff between routing fidelity and bandwidth more systematically by averaging over optimizations performed for different center frequencies. It is very clear that the desired transmission hardly depends on bandwidth whereas the logarithm of the minimal suppression of reflection and undesired transmission rises roughly exponentially with bandwidth (on average). Hence, the discrimination between desired and undesired outputs deteriorates also roughly exponentially. Nonetheless, for small bandwidths on the order of 1 MHz we can on average achieve more than 30 dB of discrimination, and for moderate bandwidths on the order of 10 MHz, we can still achieve on average almost 20 dB of discrimination, while maintaining the desired transmission above -1 dB at all frequencies within the band. These performances are certainly of technological relevance, especially in view of the extreme programmability of both the center frequency and the routing geometry. Obviously, with more elaborate optimization algorithms it may be possible to identify metasurface configurations yielding a better performance (we are unlikely to have found the global optimum), and adding more programmable meta-elements is also expected to enable better results. Nonetheless, the roughly exponential scaling of the trade-off between routing fidelity and bandwidth in our view is likely to be general. To the best of our knowledge, no bounds on this trade-off have been rigorously formulated to date; we are only aware of the well-established Bode-Fano theory[73,74] of broadband matching bounds on the suppression bandwidth of a single-port system's reflection coefficient (and its generalization to multi-port systems[75,76]). Although our goal of signal routing is more complex than mere reflection suppression, the Bode-Fano theory gives us confidence that there is indeed very likely a fundamental trade-off that could possibly be formalized via complex analysis, at least for generic coherent extinction without insisting on the maximization of desired transmission.

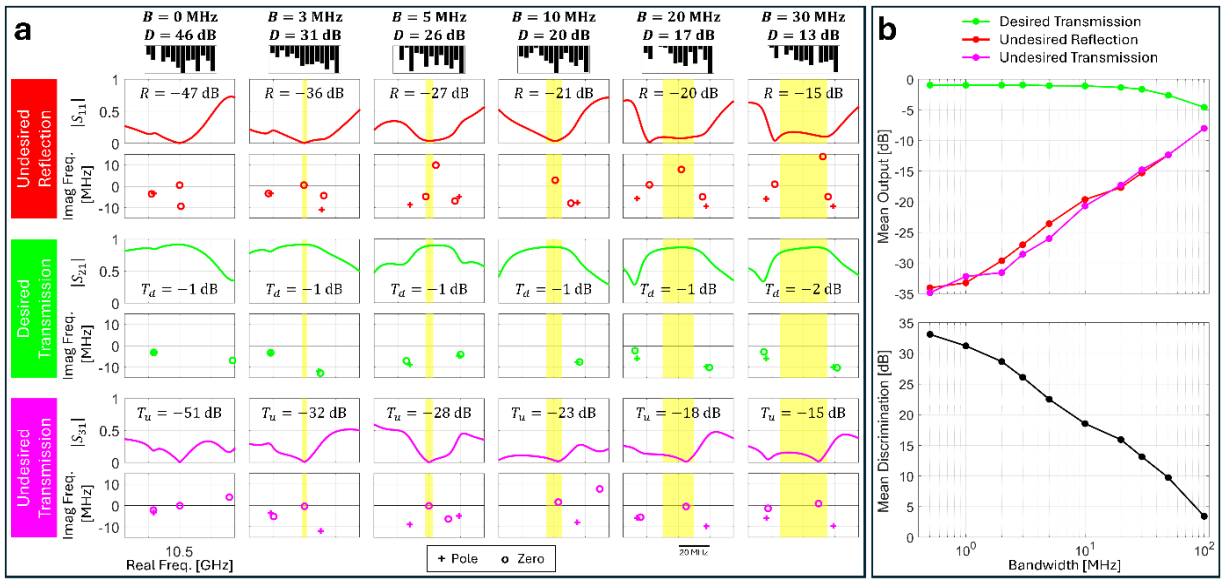

**Figure 5. Fidelity-bandwidth trade-off in (near-)reflectionless signal routing.** The same non-local programmable metasurface is reconfigured to implement a routing functionality for different choices of carrier frequency and different choices of signal bandwidth. (a) Selected examples of scattering coefficients measured with metasurface configurations optimized for a 10.5 GHz carrier frequency, for different bandwidth choices. Each column corresponds to one choice of bandwidth (highlighted in yellow) and displays the three scattering coefficients of interest: undesired reflection (top, red), desired transmission (middle, green) and undesired transmission (bottom, purple). In addition, in each case the extracted singularities (poles and zeros) are displayed. The minimal suppression or transmission is indicated in each panel. The insets on top display the metasurface configuration corresponding to each column (meta-elements sorted by index). (b) Averages over optimizations performed for different carrier frequencies of the minimal achieved suppression or transmission, and the resulting discrimination, as a function of bandwidth.

# 6. Free-Form Multi-Band Filtering in Transmission

Having investigated agile free-form filtering involving multiple scattering coefficients within one desired band in the previous sections, we now focus on multi-band free-form filtering with a single transmission coefficient. Specifically, we define multiple bands (each centered on a different center frequency, each with an individual bandwidth) and our associated filtering objectives (pass or reject). We then seek a metasurface configuration that most closely matches our filtering objectives.

We begin with the goal of rejecting signals (i.e., seeking minimal transmission) within a single band. As seen in Fig. 6a, irrespective of the chosen center frequency, we are able to identify suitable metasurface configurations that yield very good performance. For a moderate bandwidth of 10 MHz, we achieve at least 44 dB of suppression and for a very large bandwidth of 50 MHz we still achieve at least 37 dB of suppression. The scaling of the achieved suppression with bandwidth, averaged over optimizations performed for many different choices of central frequency, is roughly exponential.

Next, we test double-band objectives involving a 5 MHz pass band (i.e., seeking maximal transmission) centered on $f_1$ and a 20 MHz rejection band centered on $f_2$. As seen in Fig. 6b, for different choices of $f_1$ and $f_2$, we can simply reconfigure our metasurface and always ensure at most 1 dB attenuation within the pass band and at least 28 dB suppression in the rejection band, implying a discrimination of 27 dB.

Finally, we test triple-band filtering objectives, involving three bands of equal widths centered on $f_1$, $f_2$ and $f_3$ of which one is chosen to pass and two are chosen to reject. Again, we are able to suitably configure the metasurface in each case such that even for a large bandwidth of 20 MHz we can ensure that the desired transmission is attenuated by at most 1 dB, while the undesired transmissions are suppressed by at least 24 dB, implying a discrimination of at least 23 dB. We also once again systematically evaluated the trade-off between fidelity and bandwidth in Fig. 6c. Similar to our conclusions in Sec. 5, we observe that up to very large bandwidths it is easy to implement the pass band with minimal attenuation and that the bottleneck relates to the suppression in the rejection bands. Nonetheless, we can guarantee roughly 25 dB discrimination for a 10 MHz bandwidth on average. Combining this filtering performance with the UWB tunability and guaranteed linearity at high signal input powers, our system meets current technological needs in free-form agile signal filtering.

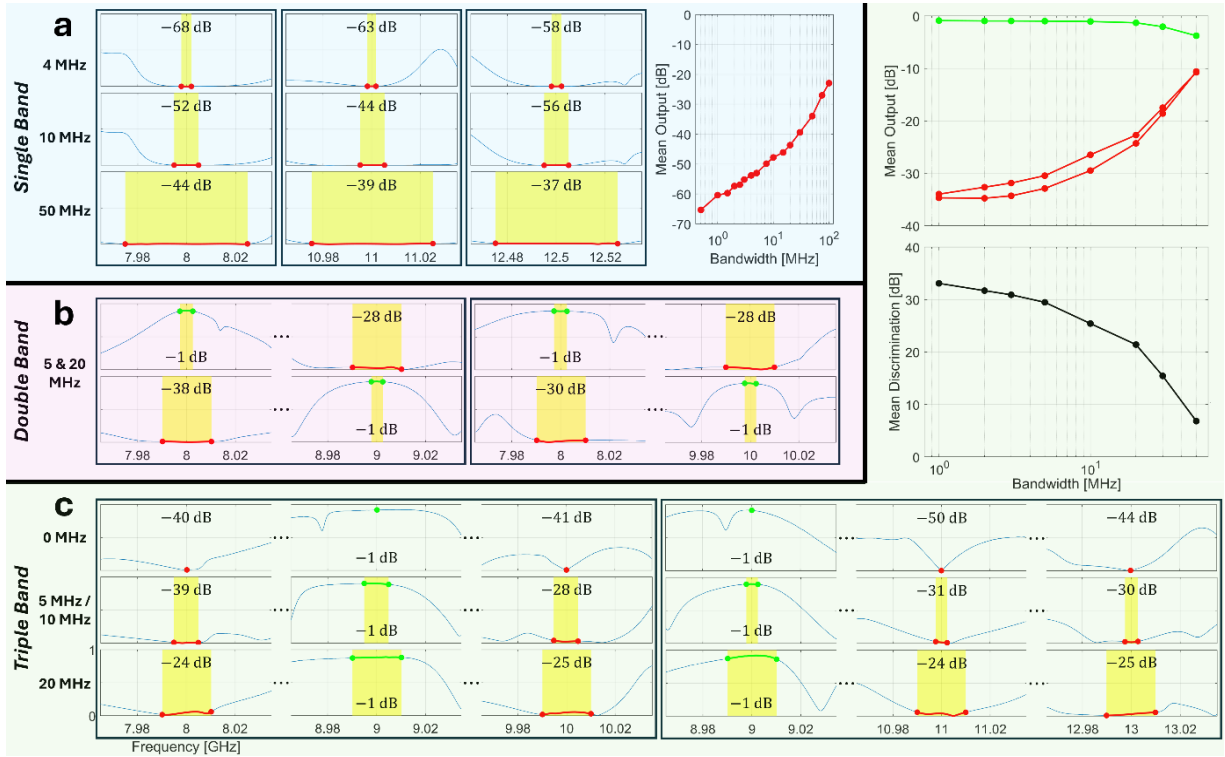

**Figure 6. Agile free-form multi-band filtering in transmission.** All displayed results are obtained by reconfiguring the same non-local programmable metasurface. (a) Selected results for single-band rejection for three different bandwidths (rows) and three different center frequencies (columns). The scaling of the minimal suppression as a function of bandwidth, averaged over different choices of central frequency, is also plotted. (b) Selected results for double-band filtering, involving a 5 MHz pass band and a 20 MHz rejection band, for four different choices of the centers of these bands. (c) Selected results for triple-band filtering, involving one pass and two rejection bands, all chosen to be of the same width. Examples for three representative choices of the bandwidth are shown, for different choices of which out of the three bands is the pass band. The scaling of the desired transmission, minimal suppressions and discrimination as a function of bandwidth, averaged over different choices of the central frequencies, is also plotted.

## 7. Summary and Conclusions

To summarize, we have introduced agile free-form signal filtering based on a reverberation-non-local programmable all-metallic metasurface device. The metasurface is backed by a chaotic low-loss cavity which provides strong all-to-all nonlocality via ergodic reverberation. Our device deliberately forgoes the limiting assumptions of conventional filter synthesis that severely constrain the design space and hence the ability to achieve UWB agile free-form filtering: our system features spatially overlapping resonances, and each meta-element's state in principle impacts all resonances and all coupling coefficients over a very large bandwidth. Using a purely optimization-based approach, we explored our virtually unconstrained design space to identify a metasurface configuration corresponding to the desired filter functionality. First, we studied the statistics of unconstrained transfer function zeros in our low-loss system, allowing us to confirm multiple recent theoretical predictions, and to demonstrate transmissionless exceptional points experimentally for the first time. Second, we demonstrated our system's control over these zeros by imposing them at any desired location on the real frequency axis within an UWB range spanning the entire X band and the lower part of the Ku band, enabling diverse modalities of meta-programmable analog signal differentiation. Third, we reported the first demonstration of low-loss reconfigurable reflectionless signal routing, where desired transmission was attenuated by as little as 1 dB, while achieving 38 dB of discrimination between desired and undesired outputs. Moreover, we could reconfigure both the operating frequency (again within an UWB range) and the choice of input and desired output ports. Fourth, we investigated the fundamental trade-off between routing functionality and bandwidth, observing a roughly exponential scaling. Nonetheless, for moderate bandwidths of 10 MHz we still achieved 20 dB of discrimination. From a theoretical point of view, we showed that the underlying optimization outcomes can be analyzed via the arrangement of the singularities (poles and various types of zeros) in the bandwidth of interest and cannot simply be understood based solely on the complex pole locations. Finally, we turned our attention to multi-band filtering in transmission, demonstrating our ability to impose multiple pass/rejection bands at desired frequencies and with desired bandwidths. Again, there is a fundamental trade-off between fidelity and bandwidth, but even for 20 MHz bandwidth we still achieved 23 dB of discrimination.

This wide range of experimental results relating to diverse modalities of entirely reconfigurable transfer function control demonstrates the ability of our device to implement agile free-form signal filtering. Due to our device's all-metallic nature, wave interactions with our device are guaranteed to be linear even at high signal powers. Moreover, we demonstrated control over the transfer function within one or multiple frequency bands at arbitrarily chosen locations within an UWB range spanning the entire X band and lower part of the Ku band. We believe that our technique would have worked even at lower frequencies than 7 GHz (but these were not accessible with our chosen coax-to-waveguide adapter due to its cut-off frequency) as long as the wavelength remains smaller than the cavity dimensions. We also believe that our setup could readily be operated at higher frequencies than 14.5 GHz provided that one is willing to tolerate the absorption which increases with frequency. Overall, the reported performances are of direct technological relevance in application domains demanding (almost) arbitrary free-form filter tunability.

Looking forward, the performance we found here can be improved in various ways. We expect to achieve open-loop control of our device by estimating the parameters of a physical model of our system (akin to Refs.[29,31]), which will dramatically enhance our ability to optimize the metasurface configuration for a given desired filter functionality. Moreover, the experimental setup can be further refined by using fast high-precision miniaturized linear actuators to implement the programmable meta-elements, as well as by using more meta-elements. The basic concept of a chaotic cavity that is reprogrammable via a mechanically actuated scattering metasurface could also be realized with a range of MEMS type hardware which may enable fast reprogramming. On the theory side, the analysis of our results has revealed that it may be possible to formulate the optimization objectives in terms of singularities (e.g., exclusion of poles from the band of interest for signal routing, and deliberate placement of various types of zeros therein). Moreover, future theoretical efforts may be able to develop useful bounds on the achievable filter performances, for example, as a function of bandwidth.

## 8. Methods

**Experimental setup.** The layout of the D-shaped chaotic cavity seen in Fig. 1a is based on a circle of 15 cm radius that is truncated on one side at a distance of 10 cm from a line crossing the circle's center and parallel to the truncated flat edge. The vertical height of the cavity is 1 cm which never exceeds half the wavelength within the considered frequency range such that the cavity is quasi-2D. The utilized motors are NEMA 8 linear stepper motors (8HY0001-T35) with 100mm T35 lead screws and controlled by a custom circuit using A4988 stepper motor drivers. The utilized coax-to-waveguide adapters are of type 16094-SF40 by Flann Microwave. The scattering matrix is measured with a vector network analyzer (Agilent Technologies PNA-L Network Analyzer N5230C) using an intermediate frequency bandwidth of 10 kHz.

**Modal overlap estimation.** According to Weyl's law[77,78], the number of modes $N_d$ below a frequency $f$ is $N_d(f) = \frac{\pi A f^2}{c^2}$ for a 2D cavity of surface area $A$, where $c$ is the speed of light. Hence, the modal density $\rho$ is $\rho(f) = \frac{\partial N_d}{\partial f} = \frac{2\pi A f}{c^2}$. The composite quality factor $Q$ being defined as $Q = \frac{f}{\Delta f}$, the mean modal overlap $\delta$ in the system is $\delta = \rho \Delta f = \frac{2\pi A f^2}{c^2 Q}$. In practice, we estimate $Q$ based on the decay rate of the impulse response envelope measured between ports and averaged across random metasurface configurations, see Ref.[55] for details. (Similar values of $\Delta f$ are also obtained via the spectral auto-correlation function.) For our cavity, $A = 0.063 \text{ m}^2$. The estimated modal overlap as a function of frequency is displayed in Fig. 1c.

**Estimation of a meta-element's effective number of independent states.** To determine the number of "independent" states of the $k$th meta-element, we fix a random configuration of the other 13 meta-elements and then sweep the $k$th meta-element through the entire range of accessible depths $c_k$ to which it is sunk into the cavity, measuring the resulting frequency-dependent scattering matrix for each $c_k$. The sweep of $c_k$ is linear and involves 300 steps. Then, for the $(i,j)$th scattering coefficient, we evaluate the effective rank[79] of the matrix $S_{ij}(f, c_k)$. Denoting by $\sigma_a$ the $a$th singular value of $S_{ij}(f, c_k)$, the effective rank is defined as $R_{ij,k}^{\text{eff}} = \exp\left(-\sum_a \sigma'_a \ln(\sigma'_a)\right)$, where $\sigma'_a = \sigma_a / (\sum_a \sigma_a)$. We repeat this procedure four times for each meta-element (i.e., the four repeats differ regarding the random configurations of the 13 other meta-elements). Then, we average $R_{ij,k}^{\text{eff}}$ across the four repeats and the six distinct scattering coefficients (three reflections and three transmissions), yielding the values of $\langle R_k^{\text{eff}} \rangle$ displayed in Fig. 1a.

**Cost function definition.** The cost function is a metric defined based on the system's scattering coefficients that is minimized by adjusting the metasurface configuration using an optimization algorithm (see below).

Fig. 3: To tune the system to feature a frequency-constrained RSM at $f_0$, the cost function is defined as the smallest eigenvalue of $\mathcal{R}(f_0)^\dagger \mathcal{R}(f_0)$, i.e., $CF = \min\left(\text{eig}\left(\mathcal{R}(f_0)^\dagger \mathcal{R}(f_0)\right)\right)$. To tune the system to simultaneously feature an RSM for different selections of ports at $f_0$, we first determine $CF_i = \min\left(\text{eig}\left(\mathcal{R}_i(f_0)^\dagger \mathcal{R}_i(f_0)\right)\right)$, where $\mathcal{R}_i$ is the reflection matrix for the $i$th choice of selected ports; then, we define $CF = \max_i \{CF_i\}$. To tune the system to feature a TSM for the transmission coefficient between the $i$th and $j$th port at $f_0$, we define $CF = |S_{ij}(f_0)|^2$.

Fig. 4a: To tune the system such that it acts as a single-frequency reflectionless signal router at $f_0$ with input port $i$, desired output port $j$, and undesired output port $k$, we first define the reflected

power $R = |S_{ii}(f_0)|^2$, the desired transmitted power $T_d = |S_{ji}(f_0)|^2$, and the undesired transmitted power $T_u = |S_{ki}(f_0)|^2$. Then, we define the cost function as $CF = \max(R, T_u, \max(\alpha - \beta T_d, 0))$, where $\alpha = 0.95$ and $\beta = 1.03$.

Fig. 5: To tune the system such that it acts as a broadband (near-)reflectionless signal router for an interval $\Delta f$ centered on the carrier frequency $f_0$, we define $CF = \max_{\Delta f}(\max(R(f), T_u(f), \max(\alpha - \beta T_d(f), 0)))$, where $\max_{\Delta f}$ denotes selecting the largest value within the interval $\Delta f$ centered on $f_0$, where $\alpha = 0.95$ and $\beta = 1.03$.

Fig. 6: We define the undesired transmitted powers $T_{u,1}$ and $T_{u,2}$ as well as the desired transmitted power $T_d$ akin to above. Then, we define our cost function as $CF = \max\left(\max_{\Delta f_1}(T_{u,1}(f)), \max_{\Delta f_2}(T_{u,2}(f)), \max_{\Delta f_3}(\max(\alpha - \beta T_d(f), 0))\right)$, where $\max_{\Delta f_i}$ denotes selecting the largest value within the interval $\Delta f_i$ centered on the $i$th carrier frequency $f_i$, and $\alpha = 0.95$ and $\beta = 1.03$.

**Optimization algorithm.** As mentioned above, we use closed-form iterative optimizations to optimize the metasurface configuration. First, we measure the scattering matrix for 100 random configurations. The random configuration yielding the lowest cost function out of the 100 considered ones is chosen as initialization for the currently best metasurface configuration. Then, we perform 200 iterations. For each iteration, we randomly pick one of the 14 meta-elements (but not the one picked in the previous iteration). Suppose we picked the $i$th meta-element. Let us denote by $c_i^{\text{curr}}$ the position of the $i$th meta-element in the currently best configuration, and the accessible positions range from 0 to $c^{\max}$. We now linearly sweep the $i$th meta-element with a step size of $\delta$ through positions spanning from $\max(0, c_i^{\text{curr}} - \Delta/2)$ to $\min(c^{\max}, c_i^{\text{curr}} + \Delta/2)$. For the first 50 iterations, $\Delta = c^{\max}/3$ and $\delta = \Delta/200$, and thereafter $\Delta = c^{\max}/6$ and $\delta = \Delta/200$. We measure the scattering matrix for each considered meta-element position; the one yielding the lowest cost function is retained to update the $i$th entry of the currently best metasurface configuration. Because of the reverberation-induced non-local interactions between the meta-elements, the scattering matrix does not depend linearly on the metasurface configuration[33]. Hence, the state of a given meta-element cannot be optimized independently of the choice of the other meta-elements' states; thus, the number of iterations strongly exceeds the number of meta-elements.

**Simulation.** The numerical simulations underlying Fig. 4b and Fig. 4c are performed with COMSOL Multiphysics. Walls and scatterer surfaces are perfect electric conductors. A plot of the real part of the out-of-plane component of the H-field is shown in Fig. 4b. The optimization is implemented by first evaluating the cost function for 1000 random configurations (orientation angles of the scatterers). The best configuration is then used as initialization in an Adam gradient-based optimization[80].

The cost function in the routing optimization in Fig. 4b is $CF = 1 - |S_{21}(f_0)|^2$. The cost function in the demultiplexing optimization in Fig. 4c is $CF = 2 - |S_{21}(f_1)|^2 - |S_{31}(f_2)|^2$. The definitions of both cost functions exploit the fact that the simulated system has zero absorption which implies flux conservation.

## Acknowledgements

P.d.H. acknowledges stimulating discussions with G. M. Rebeiz and D. R. Smith.

P.d.H. acknowledges funding from the CNRS prématuration program (project "MetaFilt"), the European Union's European Regional Development Fund, and the French region of Brittany and Rennes Métropole through the contrats de plan État-Région program (project "SOPHIE STIC&Ondes"). F.T.F. acknowledges funding from the Procope program of the French Embassy in Germany. A.D.S. and A.A. acknowledge the support of the Simons Foundation Collaboration on Extreme Wave Phenomena Based on Symmetries.

## Author Contributions

P.d.H. conceived the project. F.T.F., L.C. and P.d.H. built the experimental prototype. P.d.H. conducted the experiments, and analyzed and interpreted the data. A.A. and A.D.S. contributed to the interpretation of the experimental results and conducted the numerical simulations. P.d.H. drafted the initial manuscript. A.D.S. and P.d.H. revised the manuscript.